\documentclass[headings = small,
               fontsize = 11pt,
			   USenglish,
			   parskip = half,
			   oneside
			   ]%
               {scrartcl}

\usepackage[T1]{fontenc}
\usepackage[latin1, utf8]{inputenc}
\usepackage[USenglish]{babel}
\usepackage{csquotes}
\usepackage{bbm}
\usepackage{lmodern, microtype}
\usepackage{scrlayer-scrpage}
\usepackage{booktabs, longtable, tabularx, enumerate, array, float}
\usepackage{amsmath, amssymb, amsthm, mathtools, nicefrac}
\allowdisplaybreaks
\usepackage{graphicx, fixltx2e, colortbl, xcolor, color}
\usepackage[doublespacing]{setspace}
\usepackage{lscape}
\usepackage{comment}
\usepackage{rotating}
\usepackage{pdflscape}

\usepackage[%                           
           round,                      
           semicolon,                  
           nonamebreak,                 
           authoryear
          ]%
          {natbib}                     
\defcitealias{compustat}{S\&P Global Market Intelligence}
\defcitealias{crsp}{Center for Research in Security Prices~(CRSP)}
\defcitealias{cpi}{U.S. Bureau of Labor Statistics}
\defcitealias{factset}{FactSet}

\usepackage{abstract}
\usepackage{multicol}

\setkomafont{section}{\large\center\normalfont\scshape}
\setkomafont{subsection}{\large\center\normalfont\scshape}
\setkomafont{subsubsection}{\large\center\normalfont\scshape}

\clearscrheadings
\theoremstyle{plain}

\begin{document}
\begin{center}
\onehalfspacing 
\huge
Reference Class Selection in Similarity-Based Forecasting of Corporate Sales Growth
%vorher Titel ohne "corporate"
\footnote{Wharton Research Data Services (WRDS) was used in preparing this paper.
This service and the data available thereon constitute valuable intellectual property and trade secrets of WRDS and/or its third-party suppliers. We are very grateful for the colleagues at Flossbach von Storch (FvS) for providing the analysts' estimates and valuable advice.
This research did not receive any specific grant from funding agencies in the public, commercial, or not-for-profit sectors.}
\end{center}
\begin{center} \Large This version: \today \end{center}
%\vspace{2em}
\onehalfspacing
\parbox[t][7em][t]{0.33\textwidth}{\centering Etienne Theising\\
Institute of Econometrics and Statistics\\
University of Cologne}
\parbox[t][7em][t]{0.34\textwidth}{\centering Dominik Wied\\
Institute of Econometrics and Statistics\\
University of Cologne}
\parbox[t][7em][t]{0.33\textwidth}{\centering Daniel Ziggel\\
FOM\\
University of Applied Sciences}
%\vspace{1em}

\begin{abstract}
This paper proposes a general method to handle forecasts exposed to behavioural bias by finding appropriate outside views, in our case corporate sales forecasts of analysts. The idea is to find reference classes, i.e. peer groups, for each analyzed company separately
% gestrichen: . Hence, additional companies are considered 
that share similarities to the firm of interest with respect to a specific predictor. The classes are regarded to be optimal if the forecasted sales distributions match the actual distributions as closely as possible. The forecast quality is measured by applying goodness-of-fit tests on the estimated probability integral transformations and by comparing the predicted quantiles. The method is out-of-sample backtested on a data set consisting of 21,808 US firms over the time period 1950 - 2019, which is also descriptively analyzed. It appears that in particular the past operating margins are good predictors for the distribution of future sales. A case study compares the outside view of our distributional forecasts with actual analysts' forecasts and emphasizes the relevance of our approach in practice.
\end{abstract}

\vspace{2em}


\doublespacing
\section{Introduction}
The forecasting of future cashflows and an appropriate discount rate is pivotal for the valuation of companies and active management of equity investments \citep[e.g.][in portfolio construction]{gmx15}. In order to tackle this task, analysts have to forecast performance indicators like corporate sales or operating margins for different periods of time. However, in general there is a low predictability of growth rates \citep[see][]{chankarceskilakonishok03} and forecasts are often based on heuristics and were empirically shown to be biased as well as overoptimistic \citep[see, e.g.,][]{tvka1973, tvka1974, katv1973, cwd1988}. In our context, survey results of \cite{kunte2015} among financial market practitioners show that herding (34\%), confirmation (20\%), overconfidence (17\%), availability (15\%) and loss aversion (13\%) are the behavioral biases that affect investment decisions the most.
\cite{lim01} reviews analysts' bias, \cite{jojo12} find proof for overoptimism while \cite{loeffler98} unravels overconfidence and underreaction to news and \cite{los08} identify negligence of buisness cycles as a source of bias. 
\cite{asci07} discuss differences between buy-side and sell-side analysts' forecasts and \cite{stoni05} analyzes reasons for analysts' overconfidence.

A large part of the distorted forecasts is due to the fact that forecasts are often solely based on the so called \textit{inside view}, which considers each forecasting challenge as unique and neglects statistical information, as well as results of similar forecast challenges \citep{kalo1993}. 
Thus, it can be very helpful to use empirical data and existing experience, the so called \textit{outside view}, in order to identify and reduce the aforementioned biases \citep{tega2016}. 
The basic idea of the outside view is the definition of a reference class which includes objects of comparison similar to the initial object \citep{kahnemantversky79,loka2003}. 
By means of this objective data set the forecaster becomes empowered to challenge and improve his forecast \citep{kahnemantversky79}.
Adjusting or correcting forecasts is an already established tool in the financial and forecasting literature in terms of judgementally adjusting model based forecasts by experts \citep{woflo90,sari01,brufra17}, combining statistical forecasts with analysts' predictions \citep{lobo91,buwri91} and combining analysts' forecasts or using consensus forecasts \citep{busa99, rrs05, jjmw16}. 
However, \cite{dumc11} examine reports by research firms with multiple analysts' forecasts. Similar forecasts leads to overconfidence while highly varying forecasts diminish confidence.
Further, \cite{dubu18} show that the hit rates of analysts for earnings per share in 2014 range from 37\% to 52\%, depending on the forecast horizon.
Our contribution will add to the toolbox of analysts and investors by the property to directly calculate confidence intervals.

The concepts of the outside view and reference classes are well known in literature and practice, e.g. in infrastructure projects \citep{fl2006,fl2008,themsen19} or software development \citep{spf16}.
Moreover, the use of base rates, i.e. distributional information, is recommended by \cite{armstrong05} and is part of professional forecasters and analysts' training \citep{tega2016} which is shown to improve their performance \citep{ccmt16}.
Especially \cite{karetal21} show that the use of base rates has a positive effect on forecast accuracy but in general there has been paid more attention to the biases than to debiasing \citep{ccmt16}.
\cite{grar07} describe a procedure to include analogies in the forecasting process and \cite{lovalloclarkecamerer12} conduct an empirical study using the outside view to forecast stock returns but both suffer from a subjective choice of similar objects such that the resulting reference classes are prone to the availability bias described by \cite{tvka1973}.
Noteworthy, \cite{kkp17} construct peer groups of comparable companies for corporate valuation objectively by using a measure of similarity but these reference classes consist of only six elements elevating the probability of bias again.
Surprisingly there is a lack of studies which investigate how to construct optimal reference classes for the forecasting of future cash flows and the related performance indicators.
To the best of our knowledge, the only existing concept is proposed by \citet{maubcal15}. They define 11 reference classes based on the size of the actual sales level in order to derive base rates for the growth rate of sales. However, the defined reference classes are neither theoretically derived nor empirically backtested. Thus, the quality of the reference classes and the added value for the analysts remain vague.

%However, while the concepts of the outside view and reference classes \citep[][in infrastructure and software projects]{fl2006,fl2008,themsen19,spf16} are well known in literature and practice, surprisingly there is a lack of studies which investigate how to construct optimal reference classes for the forecasting of future cash flows and the related performance indicators. To the best of our knowledge, the only existing concept is proposed by \citet{maubcal15}. They define 11 reference classes based on the size of the actual sales level in order to derive base rates for the growth rate of sales. However, the defined reference classes are neither theoretically derived nor empirically backtested. Thus, the quality of the reference classes and the added value for the analysts remain vague.

This paper fills the previously mentioned gap in literature. On the one hand, we propose a method to find appropriate outside views for sales forecasts of analysts. Hence, we define reference classes for each analyzed company separately by means of additional companies that share similarities to the firm of interest with respect to a specific predictor. This approach is easy to implement and interpret as we deliberately restrict the analysis to exactly one predictor variable at once, which also ensures that only a parsimonious amount of data is required. Thus, the proposed method is well suited for practical applications. 
On the other hand, we evaluate different predictors and analyze their quality by means of goodness-of-fit tests and the predicted quantiles via backtesting based on a data set consisting of 21,808 US firms over the time period 1950 - 2019.
% This analysis is based on a data set consisting of 21,808 US firms over the time period 1950 - 2019 and 
This analysis yields that in particular the past operating margins are good predictors for the distribution of future sales. Moreover, in a case study we compare our forecasts with actual analysts' estimates in order to show the practical usefulness and demonstrate how to apply the results of our approach.

\section{Reference Class Selection}
The notion of reference class forecasting is based on ideas of Princeton psychologist and Nobel prize winner Daniel Kahneman and his co-author Amos Tversky. It originates in theories of planning and decision-making under uncertainties and is motivated by the fact that forecasts are often based on heuristics and were empirically shown to be biased as well as overoptimistic. In order to overcome this issue, it is advisable to contrast the inside view, i.e. information on the specific case at hand, with the outside view, i.e. information on a class of similar cases. This may include for example statistical or empirical distributional information as well as base rates and is a promising approach to overcome overoptimism, wishful thinking or strategic misrepresentations.

\cite{kahnemantversky79} introduced a corrective procedure for biases of predictions which involves five steps. First, the forecaster has to identify a set of similar cases which define the reference class and provide the distribution of outcomes to be predicted. This distribution has either to be assessed directly or to be estimated within the next step. At this point the expert uses their available information on the case for an inside prediction. In the fourth step the expert needs to assess the predictability of their forecasts. In case of linear prediction, this may be the correlation between their predictions and the outcomes. Finally, the inside prediction is corrected and adjusted towards the mean of the reference class. 

While each of the five steps has its own pitfalls in practice, we focus on the first one and provide guidance how to select an appropriate reference class. This is of major importance as \cite{kahnemantversky79} gave no guideline how to build reference classes apart from the general rule to use similar cases. Moreover, there is a fundamental conflict of objectives in defining the reference class. On the one hand, it would be desirable to take as many cases into account as possible. However, it is crucial that heterogeneity does not become too large and each object is still comparable to the initial one. On the other hand, each element within the reference class should be similar to the initial object, whereby the risk arises that the class becomes too small and the objects too similar. In this case the probability of a biased forecast is again elevated. Based on this fact \cite{loka2003} state: \textit{``Identifying the right reference class involves both art and science.''}

In literature, there are several studies dealing with reference class building. For example, \cite{lovalloclarkecamerer12} report two case studies with respect to private‐equity investment decisions and film revenue forecasts. However, and to the best of our knowledge, there is a gap with respect to reference classes for the forecasting of future cash flows and the related performance indicators. The only existing concept is proposed by \citet{maubcal15}. They state that sales growth is the most important driver of corporate value and define the reference classes by sorting the firms' real sales in 10 deciles as well as an 11th class for the top one percentile. To this end they use historical data of the S\&{}P1500 from 1994-2014. In total they show the distribution of growth rates for 55 reference classes (11 size ranges multiplied by five time horizons) but give neither a theoretical justification for nor an empirical backtest of their proposed procedure. Thus, the quality of the proposed reference classes and the added value for the analysts remain open questions, especially as they used clustered data which has a substantial problem in general. 
As an example, Figure \ref{fig:clusterbad} shows three clusters constructed by the k-means algorithm for a simulated data cloud and highlights the pitfall that an element on the border of one cluster may be closer to the elements of another cluster than to the majority of elements in its own cluster -- a general drawback of procedures using cluster algorithms. 

\begin{center}
	[insert figure \ref{fig:clusterbad} about here]
\end{center}

In order to overcome this drawback we will present an alternative method which does not rely on cluster algorithms and finds reference classes for each analyzed company separately whereby the approach is easy to implement and interpret. 
Moreover, we will evaluate the resulting reference classes out-of-sample on a 1950--2019 data set in order to be able to make a meaningful quality valuation. The following two subsections will provide the theoretical foundations.

\subsection{Theoretical Framework}
We aim to forecast $Y_{i,t+h}$, i.e. an $h$-step ahead forecast of the random variable $\{Y_{i,t}\}$ for firm $i$ at time $t$. In the following applications this will be the sales growth but basically it could be any other quantity of interest. At this point we assume that a sufficient amount of historical data of additional firms is available in order to assess the distribution of $Y_{i,t+h}$. We base the reference class on a specific reference characteristic $\{X_{i, t}\}$.\footnote{For sake of readability we have restricted the notation in such a way that the subsequent applications are covered. In principle, the model also allows for several reference characteristics with time series properties.} 
The idea is now to build a reference class $J$ by finding firms $j$ in the past which are similar to firm $i$ with respect to the reference characteristic and in some norm $\vert \vert \cdot \vert \vert$, i.e.
\begin{align*}
	\vert \vert \{X_{i, t}\} - \{X_{j, s}\} \vert \vert
\end{align*}

shall be \textit{small}, where $s + h\leq t$ to ensure the realization of $Y_{j,s+h}$ is available. 
For example, we could use all companies which had an operating margin $\pm 1$ percentage points in comparison to the actual margin of firm $i$ during the last 10 years. Figure \ref{fig:clusterbad} illustrates the difference of our approach to a classical cluster analysis. We do not try to find disjoint clusters of firms, but aim at finding neighbors for each firm separately. A forecast for the distribution of $Y_{i,t+h}$, which is used as an outside view, is now given by the empirical distribution of the values $Y_{j,s+h}$, $(j,s)\in J$. 

The first assumption behind the approach is the existence of a \textit{market mechanism}, say a smooth function $f_h$ such that $Y_{i,t+h}\sim f_h(\{X_{i, t}\})$. Moreover, we need some kind of stationarity assumption so that this mechanism works similarly over time and we have $Y_{j,s+h} \sim f_h(\{X_{j, s}\})$, $(j,s)\in J$, for the outcomes within the reference class. If $\{X_{i, t}\}$ is close to $\{X_{j, s}\}$, which is supposed to be provided by finding suitable reference classes, $f_h(\{X_{i, t}\})$ is close to $f_h(\{X_{j, s}\})$ and the empirical distribution function of $Y_{j,s+h}$ is a good approximation for the distribution of $Y_{i,t+h}$. Note, the goal of this paper is not to get information about $f_h$, but to get information about how suitable reference classes are.

%If further $\vert \vert \{X_{i,\tau}\}_{t^* \leq \tau \leq t} - \{X_{j,\tau}\}_{s^* \leq \tau \leq s} \vert \vert < \delta$ and $f$ is continuous, then $\vert\vert f(\{X_{i,\tau}\}_{t^* \leq \tau \leq t}) - f(\{X_{j,\tau}\}_{s^* \leq \tau \leq s})\vert\vert^* < \varepsilon$ in a suitable norm $\vert\vert \cdot \vert\vert^*$, and the ecdf of $Y_{j,s+h}$ approximates the distribution of $Y_{i,t+h}$.

\subsection{Performance of Procedure}
By means of the resulting distributional information we can assess predictions (e.g. by experts or analysts or model based forecasts) or we can assess the suitability of the reference class by evaluating the empirical cumulative distribution function of the reference class at the (known) realization, i.e. we calculate
\begin{align}\label{PIT}
	\mathbbm{P} (Y_{i,t+h} \leq y_{i,t+h}) \approx n^{-1} \sum_{(j,s)\in J} \mathbbm{1}\{ Y_{j,s+h} \leq y_{i,t+h}\},
\end{align}
where $n=\vert J\vert$. Repeating this for multiple firms and points in time results in a sample of size $m$, whereas the values lie in the interval $[0,1]$. 
If the approximation of the distribution is valid, \eqref{PIT} is roughly the probability integral transform and consequently we approximately have realizations from a uniform distribution on $[0,1]$. 
To assess the forecast ability of the different predictor variables, we consider measures that determine how close this approximation is. This is done with classical statistical goodness-of-fit tests as well as a comparison of quantiles.

Let $F_m$ be the empirical distribution function of these frequencies $\{p_k\}_{k=1,\dots, m}$ and let $F$ be the true distribution function of the counterparts of these frequencies in the population. 
Let $F_0$ be the distribution function of the uniform distribution on $[0,1]$. The considered hypothesis pair is $H_0:\; F = F_0$ vs. $H_1:\; F \neq F_0$ and the corresponding two test statistics are given by $\sqrt{m} \sup_{x\in [0,1]} \vert F_m (x) - F_0 (x)\vert$ (Kolmogorov-Smirnov) and $m\int_{0}^{1} [F_m(x)-F_0(x)]^2\mathrm{d}F_0(x)$ (Cramer-von-Mises).

However, we do not consider the actual tests' decisions. Working with sample sizes between $100,000$ and $300,000$, depending on hyper parameters, we face the problem pointed out by \cite{berkson38}: ``Any consistent test will detect any arbitrary small change in the [distribution] if the sample size is sufficiently large''. Thus, most p-values would be very small or even get reported as $0$ by software. Avoiding this problem, we focus on the value of the test statistics, i.e. we rank the different combinations of predictor variable and hyper parameters based on these values.

%Draw subsamples and evaluate the distribution of p-values.
%Perhaps, use a test for relevant distributional distance.

A third measure of ranking the models consists of comparing the quantiles. This means that for a finite number of quantile levels, we consider the absolute difference between the quantiles of $\{p_k\}_{k=1,\dots, m}$ and the quantiles of the uniform distribution on $[0,1]$. These differences are summed up and ranked.

% consider the absolute difference between the forecasted sales growth quantiles and the observed quantiles.

%Anderson-Darling test: $m\int_{0}^{1} \frac{[F_m(x)-F_0(x)]^2}{F_0(x)(1-F_0(x))}\mathrm{d}F_0(x)$.

\begin{comment}
\subsubsection{Problems of AD-Test}
Most algorithms use the test statistic
\begin{align*}
	-m - m^{-1}\sum_{i=1}^m (2i-1)\ln (p_i(1-p_i))
\end{align*}
which is not defined if there is any $i$ with $p_i=0$ or $p_i=1$.
It is possible to incorporate $p_i=1$ by some algebra but for any observation $i$ with $p_i=0$ the integral 
\begin{align*}
	\int_{0}^{1} \frac{[F_m(x)-F_0(x)]^2}{F_0(x)(1-F_0(x))}\mathrm{d}F_0(x)
\end{align*}
is not well defined.
\end{comment}

\section{Data Set}
In order to find the best predictor variable and appropriate hyper parameters we analyze their 
%historic 
performance on an historic data set 
with regards to finding optimal reference classes. We use Compustat North America fundamentals annual data\footnote{Downloaded 28 January 2020} from 1950 to 2019 by \citet{compustat} and limit our analysis to US firms excluding companies from the financial and real-estate sector.
Firms without sales information or only one observation are discarded due to our interest in predicting distributions of sales growth.
We merge these data with stock-exchange information from the \citet[CRSP,][]{crsp} daily stock\footnote{Downloaded 30 January 2020} of the University of Chicago Booth School of Business.
All variables collected in US dollar are inflation adjusted to 1982 -- 1984 US dollar using monthly inflation rate data from the consumer price index for all urban consumers\footnote{Downloaded 23 January 2020} (all items in US city average) by the \citet{cpi}.

The data set consists of 303,628 observations on 21,808 firms with CRSP stock exchange market information on 206,221 observations of 17,099 firms in total.
The length of the time series of the different firms varies considerably (c.f. Figures \ref{fig:tslengthbarplot} and \ref{fig:tslength}) as well as the number of observations per year (c.f. Figure \ref{fig:obsperyear}).
To put this in perspective, there is an influence of survivorship in the data set.
Our later backtest focusses on one, three, five and 10 year predictions and the survivorship rates are 97.25\% for one year, 89.61\% for three years, 76.12\% for five years and 48.20\% for 10 years.

\begin{center}
	[insert figures \ref{fig:obsperyear}, \ref{fig:tslengthbarplot} and \ref{fig:tslength} about here]
\end{center}

We select and investigate the most common metrics used for fundamental analysis as possible predictor variables whereby some of them relate to the company directly while some others are market parameters. To be more precise, observed key figures for all companies are sales, operating margin, total assets, shareholder equity, the SIC (standard industrial classification), $\beta$, the price-to-earnings ratio and the price-to-book ratio.
Using sales and operating margin information over time, we construct one to 10 year past sales growth and one to 10 year past operating margin delta as additional possible predictor variables where the necessary data are available.
Instead of SIC itself, we derive a firm's major and industry group and use these groups to construct reference classes as a benchmark of the typical current practice.
In Table \ref{tab:predictors} we provide a summary of the predictor variables used to construct reference classes including a description, relevant quantiles, their means and the number of missing values in the data set.

\begin{center}
	[insert tables \ref{tab:predictors} and \ref{tab:cagr-fulluniverse} about here]
\end{center}

We aim to forecast distributions of future sales growth while using exactly one of the predictor variables to construct reference classes.
To be more precise, we construct one, three, five and 10 year future sales growth forecasts using temporal information in the data set.
Table \ref{tab:cagr-fulluniverse} displays the base rates, i.e. the historical sales compound annual growth rate (CAGR), for the full universe of data.
Here, the tails of the distribution get lighter, the (2.5\%-trimmed) standard deviation declines, the (2.5\%-trimmed) mean gets closer to the median and the distribution more centered the longer the forecast horizon is, as it is visible in Figure \ref{fig:cagr-dens} as well.
By a 2.5\%-trimmed mean or standard deviation we are referring to the arithmetic mean or standard deviation, respectively, where the largest 2.5\% and the smallest 2.5\% of the data are excluded.\footnote{To be precise, for a vector of sorted observations $\{x_i\}_{i=1,\dots, n}$ we compute any $\alpha$-trimmed measure, $0< \alpha <1$, based on the trimmed vector of observations $\{x_i\}_{i=[\alpha n]+1,\dots , n - [\alpha n]}$, where $[\cdot ]$ is the floor function.}
The (2.5\%-trimmed) means of sales CAGR are larger than the respective medians because the growth rates are left bounded and right unbounded and we observe a substantial amount of high values one could characterize as outliers which make the ordinary mean and standard deviation uninformative.
In order to restrain the influence of these outliers and to keep the mean and standard deviation informative we use the trimmed versions of these measures.
The summary statistic of the sales CAGR can be found in Table \ref{tab:predictors} as the distribution of future and past growth rates in the full data set are identical.

\begin{center}
	[insert figure \ref{fig:cagr-dens} about here]
\end{center}

\section{Backtest}
By means of a backtest we compare the performance of our new procedure to forecast distributions of sales growth rates to the performance of the benchmark approach by \cite{maubcal15} and the typical practice of using industry classifications, here the first two and three digits of SIC, respectively.
We include three (hyper) parameters in the backtest where all methods depend on the number of past years to use for reference class construction and only our new procedure depends additionally on the predictor variable as well as the size of the reference class (see Table \ref{tab:parameters}).
Forecast horizons investigated are one, three, five and 10 years.
%We aim to forecast sales growth rates for different time horizons $h$ and to use only contemporaneous information, as well as a single predictor variable in order to build the reference classes. Besides a single predictor variable we consider two further (hyper) parameters of our procedure as part of the case study (see Table \ref{tab:parameters}). 

\begin{center}
	[insert table \ref{tab:parameters} about here]
\end{center}

The parameter window $w$ defines the number of past years to provide candidates of historical observations to construct a reference class.
All observations from this window period with known outcomes, i.e. firms with available $h$-year future sales growth, are candidates for the reference class.
In order to backtest out-of-sample, given an initial case firm $i$ at time $t$, the parameters $w$ and $h$ determine the years of historical data to serve as candidates, namely starting in $t - h - w + 1$ and ending in $t - h$ (assuming that at time $t$ all information of the financial year $t$ is available).
That means we consider all firms $j$ at times $s$ as candidates for the initial case's reference class, where $t-h-w+1 \leq s \leq t-h$ and the predictor variable and $h$-year sales growth are available.

The size of the reference class, i.e.\ the number of observations it contains, is relative to the number of candidates and defined by the size parameter $c \in (0,1)$ determining which  of the candidates $X_{j,s}$ lie closely enough to the initial case $X_{i,t}$ to be a member of the reference class. 
To be more precise, this means $c$ assesses for which candidate firms $j$ at time $s$ the value $\vert\vert X_{i,t} - X_{j,s}\vert\vert$ is considered as \textit{small}.
Here, we order the candidates by the predictor variable and take the $c/2$ fraction smaller than the initial case's observation and the $c/2$ fraction larger than the initial case's observation.
More theoretically, let $\hat{F}_{\text{cand}}$ be the empirical distribution function of all candidates and $\hat{F}^{-1}_{\text{cand}}$ be the associated empirical quantile function of all candidates.
Then, all candidates $\{j,s\}$, i.e. firms $j$ at time $s$, with $\vert \hat{F}^{-1}_{\text{cand}}(X_{i,t}) - \hat{F}^{-1}_{\text{cand}}(X_{j,s})\vert \leq c/2$ are chosen as members of the reference class.
The parameter $c$ is only relevant for our new approach.
To keep the class size constant even if the initial case's predictor variable is at the tail of the candidates' distribution, we choose the top or bottom fraction $c$ of the candidates regarding the predictor variable if $\hat{F}^{-1}_{\text{cand}}(X_{i,t}) > 1 - c/2$ or $\hat{F}^{-1}_{\text{cand}}(X_{i,t}) < c/2$, respectively. Moreover, the reference class of each case has to consist of at least 20 elements or members in order to allow reasonable distribution forecasts and to be considered within our backtest, this requirement applies to the benchmark methods as well.

The benchmarks models are the approach of \cite{maubcal15} and a simple approach using the major and industry group of a firm and set the bar for our new method.
\cite{maubcal15} define the reference classes by sorting the candidates' real sales in 10 deciles as well as an 11th class for the top one percentile.
We use the major and the industry group in a typical straightforward way to construct a reference class from the set of candidates.
In both cases, all candidate firms that are in the same major or industry group, respectively, as the initial case are members of the reference class.
Thus, there is no size parameter in either of the benchmark approaches.

Our new approach is analyzed with regards to 27 predictor variables, three different class sizes and four different window lengths, thus resulting in 324 different combinations for each forecast horizon.
The approach of \cite{maubcal15} uses one predictor variable and four different window sizes, i.e. four combinations for each forecast horizon, and the typical industry classification approach uses two predictor variables and four different window sizes, i.e. eight combinations.
In total we have 336 different combinations for each forecast horizon.

For each approach and combination of (hyper) parameters we consider each observation in the data set, i.e. each firm $i$ at each point in time $t$ (where the firm is in the data set), as an initial case.
We construct a reference class if several criteria are met.
The predictor variable and the full window length of historical data must be available, i.e. $t \geq 1950 + w + h - 1$ since our data set starts in 1950.
The $h$-year future sales growth must be available, so at least $t \leq 2019 - h$. Moreover, firm $i$ must be in the data set at time $t + h$ and the reference class has to consist of at least 20 elements.

After obtaining the reference class for an initial case $(i,t)$ we evaluate the empirical distribution function of the sales growth rates of the reference class elements (base rates) at the realized sales growth rate of firm $i$ at time $t$.
Doing this for all initial cases of a parameter combination provides a sample of forecasted probabilities $\{p_k\}_{k=1,\dots, m}$ of being less or equal to the realized sales growth of the initial case. 
%of size $m$ depending on the availability of the predictor and forecast variable, the window length and the forecast horizon.
The sample size $m$ depends on the availability of the predictor and forecast variable, the window length and the forecast horizon.
If the approximation of the distribution by the reference class is valid we roughly have realizations from a uniform distribution on $[0, 1]$.
We then use the Kolmogorov-Smirnov (KS) test statistic and the Cramer-von-Mises (CvM) test statistic to measure the accuracy of the distributional approximation.
As a third measure of the accuracy, we calculate the differences of the 1\%, 5\%, 10\%, 25\%, 50\%, 75\%, 90\%, 95\% and 99\% quantiles of $\{p_k\}_{k=1,\dots, m}$ and of the uniform distribution on $[0,1]$, respectively, and sum up the absolute values of these differences ($\Delta_{\text{quantiles}}$).

\subsection{Results of Backtest}

Tables \ref{tab:result1yearsalesgr} - \ref{tab:result10yearsalesgr} show an excerpt of our results\footnote{Full results are available upon request.}.
We display the best three parameter combinations according to the quantile deviation $\Delta_{\text{quantiles}}$ and as a comparison the benchmark approach of \cite{maubcal15} for the best window length. 
Moreover, we present the benchmark approaches using industry classification through SIC's major and industry group with the best window length, respectively.
The best combinations are in all cases various combinations of the predictor \textit{past operating margin delta} followed next by the predictor \textit{operating margin} which is why we included the best parameter combination for the operating margin as well.
As a comparison to the simpler approach by \cite{maubcal15} we also included the best parameter combination for the predictor \textit{sales}.
All predictor variables which include only contemporaneous information have the common advantage not to rely on (a lot) of historical information of the initial case.\footnote{The necessity of historical information to use the past operating margin deltas as predictors reduces the amount of data and produces the risk of survivorship bias causing the better accuracy. We performed a robustness check where we limited the data set for each forecast horizon to the observations with available best predictor variable of this backtest. 
The past operating margin deltas still performed best. Results are available upon request.}
The best parameter combinations all involve a window length of 30 which may be hard to achieve in practice.
Hence, we added the best parameter combinations for window lengths five and 10 to get an impression of the influence of historical information. 
Thus, we report 10 results for each forecast horizon except for one-year sales growth. 
Here, the best parameter combination for window length 10 and the best parameter combination for predictor \textit{operating margin} coincide.

In order to get a sense of the measure $\Delta_{\text{quantiles}}$, we consider the best predictor \textit{six-year operating margin delta} for forecasting one-year ahead sales growth from Table \ref{tab:result1yearsalesgr}.
Here we have $\Delta_{\text{quantiles}} = 0.0155$, which is the sum of the absolute quantile deviations for nine quantiles. So, the mean absolute deviation of these quantiles is $0.17$ percentage points.
Therefore, the backtest shows that we miss the quantile levels of the underlying distribution of one-year ahead sales growth on historical data by only $0.17$ percentage points on average. Assuming e.g. that a practitioner constructs a $95\%$ confidence interval from the reference class the error in coverage rate should be negligible.

\begin{center}
	[insert tables \ref{tab:result1yearsalesgr}, \ref{tab:result3yearsalesgr}, \ref{tab:result5yearsalesgr} and \ref{tab:result10yearsalesgr} about here]
\end{center}

The results are consistent across the accuracy measures and the relative class size does not influence the results substantially.
All goodness-of-fit measures generally improve with a shorter forecast horizon.
The \textit{past operating margin deltas} are the best predictor variables using a window of length 30. In contrast, the best predictor variables for window lengths of five and 10 are the \textit{operating margin} for forecast horizons one and three while the \textit{price-to-earnings ratio} is best for the forecast horizon five. For forecast horizon 10 \textit{price-to-earnings ratio} is optimal for the window length five and the \textit{10-year past operating margin delta} for a window length of 10.

Constructing reference classes by the benchmark procedure using \textit{major or industry groups} yields the worst results for horizons one, three and five. 
Only for a 10-year horizon the industry classification by groups results in more accurate distributional forecasts.
The approach by \cite{maubcal15} performs in a very similar way to using \textit{sales} as a predictor in our approach.
For forecast horizons one, three and five their approach is slightly better than ours using \textit{sales} and for a 10-year horizon it is vice versa.
Nonetheless, their approach performs clearly worse than the best parameter combinations according to our accuracy measures.

Although it is not the aim of this work to give a theoretical framework of the drivers of sales growth, we will try to give some intuition behind the results presented above, especially as the operating margin or its past delta are not commonly known as drivers of sales growth. 
Both figures are cumulative metrics which condense a lot of information. 
For example, the competition within the industry \citep[see, e.g.,][]{porter1979} or the competitive position of the company \citep[see, e.g.,][]{porter1985} significantly affect the operating margin (deltas) as well as the future development of a company.
Intuitively, the more a company's operating margin grows the better is its market position and it is natural to expect a higher sales growth.
This corresponds to the results in Table \ref{tab:pred-infl} discussed below.
Thus, it is not too surprising that the predictor variables \textit{operating margin} and \textit{past operating margin deltas} perform better than other variables including much less information. 
With respect to the benchmark approach of \cite{maubcal15} the superior performance could be partly explained by Gibrat's law which basically states that the proportional rate of growth of a company is independent of the absolute size \citep{gibrat1931}.   

To get a feeling for the influence of the predictor variable in our new approach on the shape of the distribution forecast provided by the reference class, we consider the year 2018 as an example in view of the later application in practice.
For each forecast horizon we use the best parameter combination, according to the measure of quantile deviations $\Delta_{\text{quantiles}}$ and construct artificial initial cases by calculating the 10\% to 90\% quantiles of the predictor variable. After that, we use our new approach to construct reference classes based on these initial cases.
Table \ref{tab:pred-infl} displays the value of the predictor variables and the median, mean and standard deviation of the distributional forecast of the associated quantiles.

\begin{center}
	[insert Table \ref{tab:pred-infl} about here]
\end{center}

The location and scale parameters behave similarly for all forecasting horizons. 
The standard deviation is smallest for medial predictor variables and rises towards the tails reflecting the uncertainty in the tails of the distributions by this v-shape. The mean and median are monotone in the predictor quantiles besides few exceptions indicating that higher past margin deltas coincide with higher sales growth. 

\section{The Outside View in Practice}

In the last section we systematically investigated the accuracy of constructing reference classes using a single predictor variable.
In practice, we are able to assess a prediction by evaluating the empirical distribution function of the reference class. Thus, we can use the distributional information, i.e. the outside view, of the reference class to correct a potentially flawed or biased prediction. Moreover, we can calculate point forecasts
%estimates 
based on the median or mean of the reference class, confidence intervals based on the quantiles of the distributional forecast, or similarity-based forecasts using the outcomes of the reference class and weighting them according to a measure of similarity to the initial case. 

However, in order to demonstrate how to use our method in practice, we compare the resulting outside view with experts' forecasts
and calculate base rates for two examples -- 3M and Amazon. To be more precise, for both companies we forecast the distribution of one-year annual sales growth based on the best combination of predictor variable and hyper parameters. 
These results are compared to analysts' forecasts which were obtained from the \citet{factset} estimates database\footnote{Downloaded 07 January 2021}, whereby for both estimates 2018 is the base year.\footnote{We also calculated the distribution for the three-year sales growth but the results are very similar with respect to the basic statement, thus we only report the one-year results. Moreover, we could not take longer prediction horizons into account as there were far too few observations available.} The results are presented in Figures \ref{fig:example-dreim-1} and \ref{fig:example-amazon-1}.

\begin{center}
	[insert figures \ref{fig:example-dreim-1} and \ref{fig:example-amazon-1} about here]
\end{center}

For 3M there are 15 expert forecasts and Figure \ref{fig:example-dreim-1} illustrates that these forecasts vary between -2.35\% and 3.26\% and lie slightly below the median of our forecasted distribution. Thus, there is no indication of overoptimistic forecasts as in- and outside views coincide. Both views classify 3M as an average company with respect to sales growth. However, the low variability of forecasts may lead investors to overconfidence in the reported range of forecasts. The outside view uncovers higher sales growth varibility, thus preventing the overconfidence pitfall.

Figure \ref{fig:example-amazon-1} shows the results for Amazon, based on 43 expert forecasts, which differ considerably. On the one hand, the forecasts are more heterogeneous and vary between 13.93\% and 22.82\%. On the other hand, the forecasts are much more optimistic and correspond to quantiles between 76.87\% and 88.25\%. 
This means that for the most optimistic forecast, roughly only one out of 10 companies within the reference class managed to reach the forecasted growth of Amazon. 
This big difference between in- and outside views should at least exhort the analysts to scrutinize their forecasts and to question the arguments for the optimistic assessment. Although Amazon is well known to be a high-growth company the analysts should have good reasons for such optimistic forecasts.

\begin{center}
	[insert tables \ref{tab:refclass_dreim} and \ref{tab:refclass_amazon} about here]
\end{center}

Tables \ref{tab:refclass_dreim} and \ref{tab:refclass_amazon} are inspired by \cite{maubcal15} and show the base rates for 3M and Amazon. At this point it is worthwhile mentioning that our method yields different base rates for each company while the method of Mauboussin and Callahan results only in 11 clusters with one set of base rates for each. Furthermore, it is noteworthy that for both companies, and every forecast horizon, the mean, median as well as standard deviation are higher for our reference classes. This is due to the fact that small firms are included within our reference classes. This observation is in line with the results presented in \cite{maubcal15} where these figures also increase with decreasing sizes of companies. As 3M and Amazon are relatively large companies with sales of USD 32.7 and 232.9 billion in 2018, respectively, small companies are not included in the reference classes of Mauboussin and Callahan. As a further consequence, the base rates of our approach are less concentrated in the range -5\% to 10\% and imply a wider range of possible outcomes which appears realistic. However, we do not want to make an assessment of the procedures as this point as this was already done within the last section.

\section{Conclusion and Outlook}
%All goodness of fit measures rise with larger forecast horizon.
%Past operating margin deltas and the operating margin are the best single predictor variables.
%Larger windows of historical data yield better results but the class size is negligible, 2.5\% and 5\% are among the best in all forecast horizons.
In this paper, we have extended financial analysts and investors' toolbox by a general method to provide outside views for forecasting sales growth and we have provided an extensive backtest on sales data from the USA over several decades. Additionally, we have compared the method to several benchmark approaches used in practice and applied it to real world examples of 3M and Amazon. The new approach delivers very reasonable results, needs only a parsimonious amount of data and is easy to interpret. Thus, it is well suited to applications in practice and lays a sound foundation for further research as several extensions of our approach are possible.

First, the method itself can be extended by including multiple predictor variables or time series characteristics. In our approach, we focus on the case of one variable having an easy interpretation and a direct extension of the approach by \citet{maubcal15} in mind. Clearly, it would be interesting to see if better reference classes could be constructed with more than one predictor variable.

Within our method, the crucial part is to find orderings of the forecast ability of the different predictor variables based on several quality criteria. We have not answered the question in which sense the different forecasts are statistically significantly different. Moreover, it is still an open question which forecast variables are actually acceptable for generating appropriate outside views and which not, i.e. it would be interesting to know in which numerical regions the goodness-of-fit measures may or may not lie. Maybe, a testing approach for relevant differences like \citet{dettewied:2016} could be helpful here. The thresholds could be determined by potential losses induced by correcting the experts' forecasts \citep[which][proposes]{kahnemantversky79}, for example.

Finally, several stress tests of our method are possible. One could perform a simulation study to assess how well reference classes can uncover true underlying distributions of any variable in order to better understand the mechanics of reference classes. Furthermore, a formal approach of correcting potentially biased expert forecasts with the similarity-based outside views can be worked out and backtested. This means that one would consider point forecasts based on the median or mean of the forecasted distributions, combine them suitably with the experts' views and backtest whether these combinations lead to better overall forecasts.

%similartiy-based forecasts can be constructed using the outcomes of the reference class and weighting them according to a measure of similarity to the initial case. 

\section*{Disclosure Statement}
The authors report there are no competing interests to declare.

\bibliography{LiteratureReferenceClassSelection}

\appendix
\section{Figures and Tables}

\begin{minipage}[c][\textheight][c]{\textwidth}
\begin{figure}[H]
	\center
	\includegraphics[height=0.33\textheight]{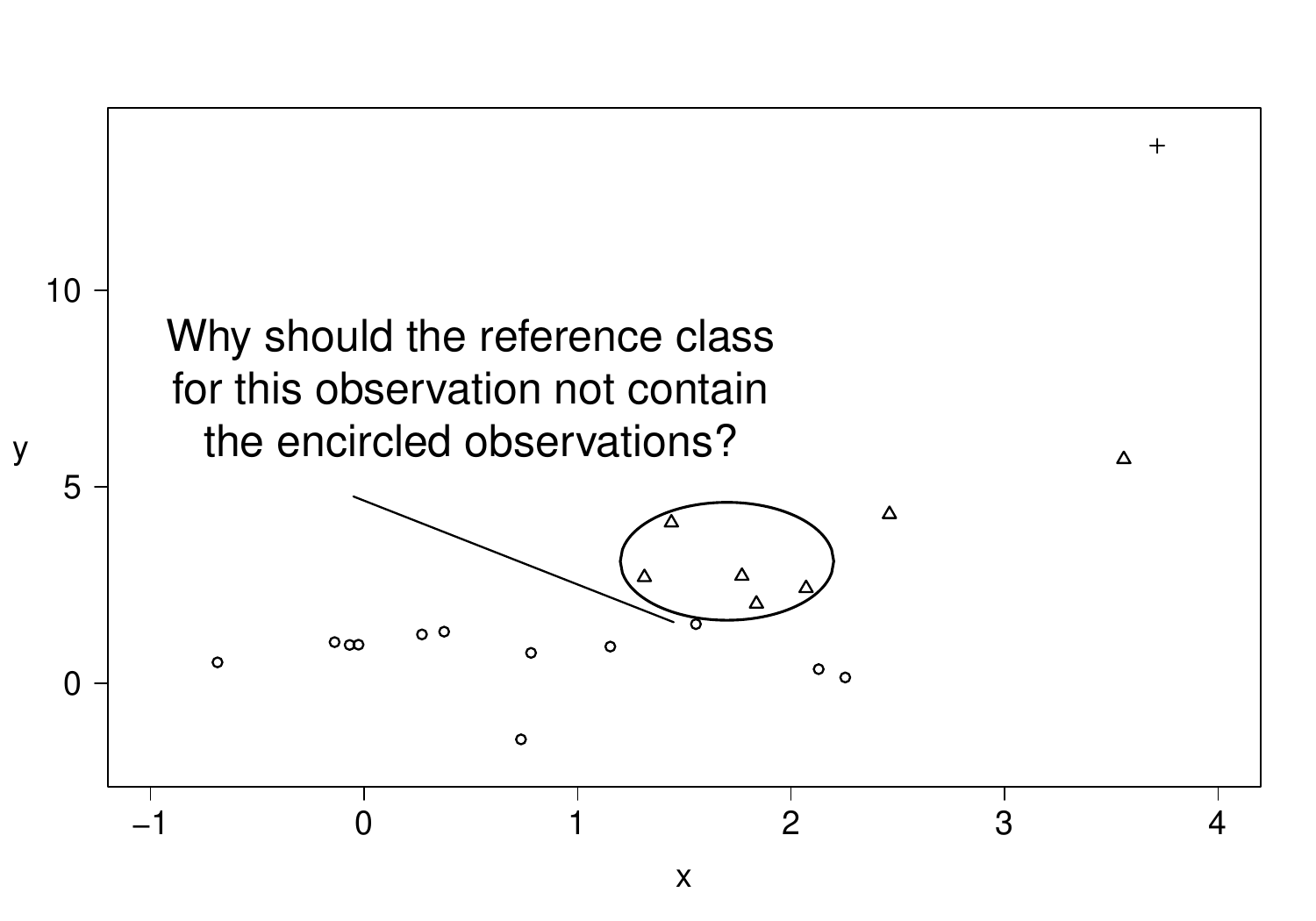}
	\caption{These three clusters constructed by the k-means algorithm for a simulated data cloud highlight the pitfall that elements on the border of one cluster may be closer to the elements of another cluster than to the majority of elements in their own clusters. By not building clusters but custom reference classes for each forecasting instance we overcome this disadvantage.}
	\label{fig:clusterbad}
\end{figure}
%alte caption: Three clusters by the k-means algorithm for an artificial data cloud.

\vspace{2em}

\begin{figure}[H]
	\center
	\includegraphics[height=0.33\textheight]{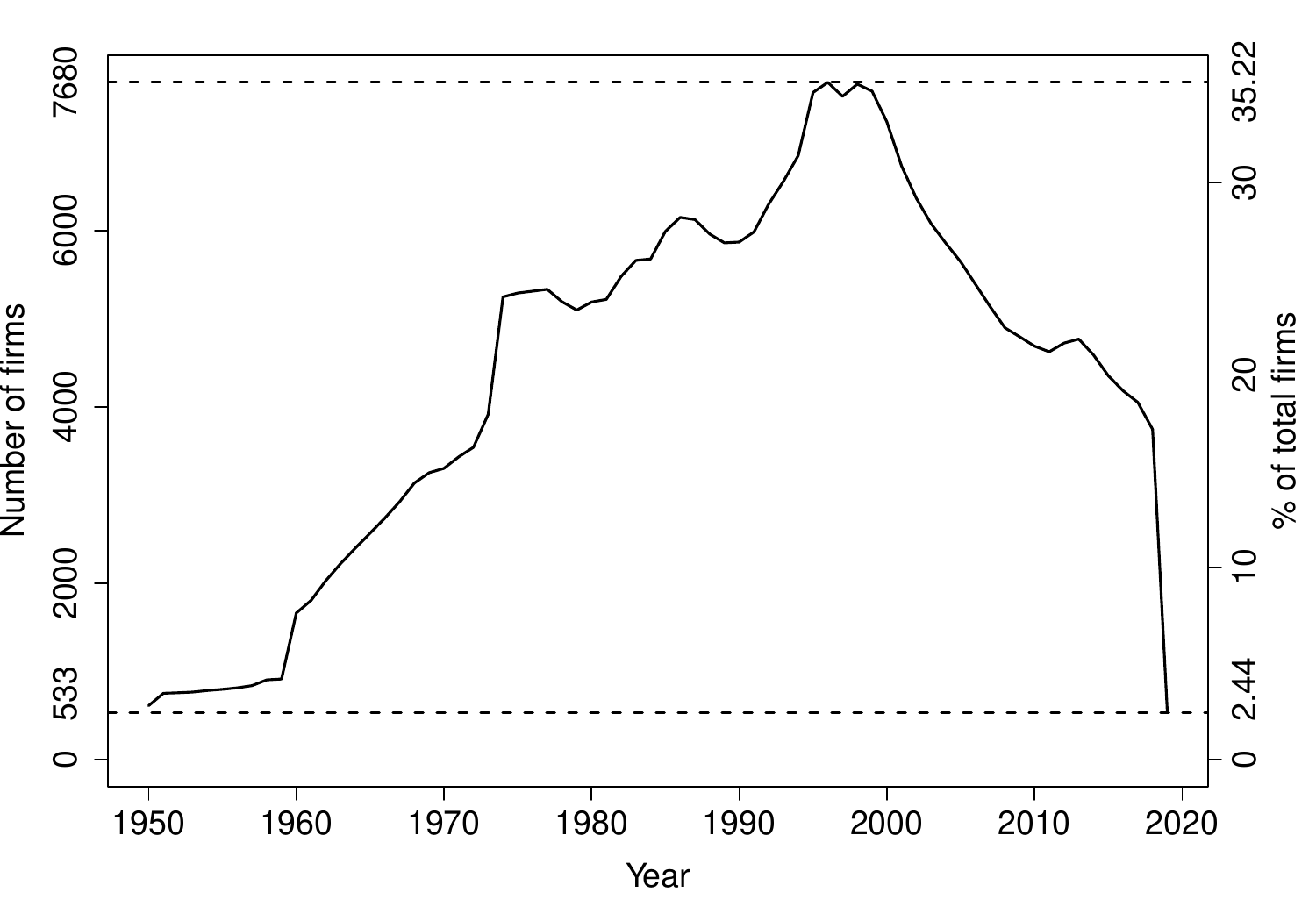}
	\caption{Number of companies over time. The left vertical axis covers the number of firms, i.e. observations, per year and the right vertical axis covers the number of firms as a proportion of the total number of firms.}
	\label{fig:obsperyear}
\end{figure}
\end{minipage}

\begin{minipage}[c][\textheight][c]{\textwidth}
\begin{figure}[H]
	\center
	\includegraphics[height=0.33\textheight]{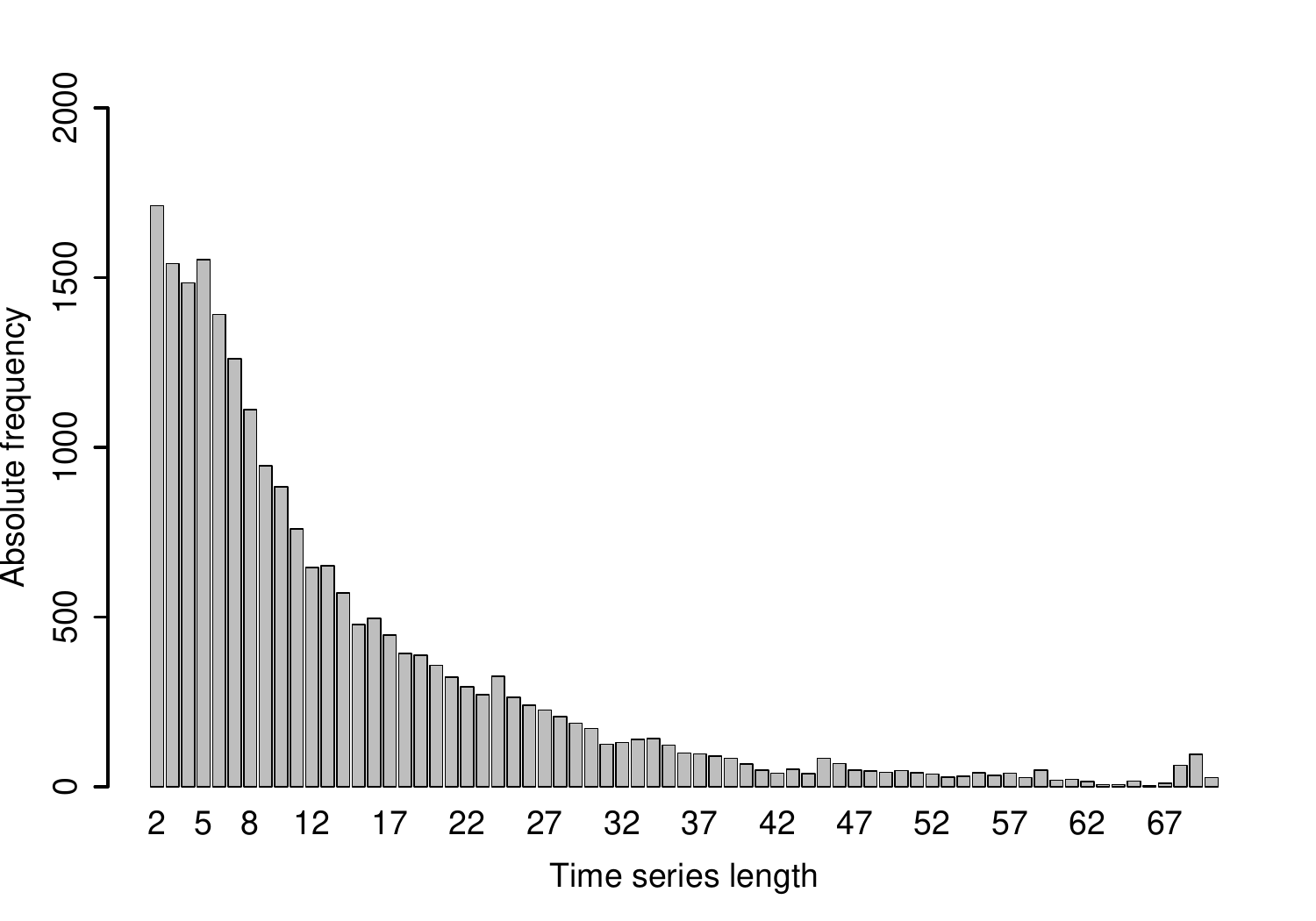}
	\caption{Barplot of observations per firm in the data set displaying the empirical distribution of time series length.}
	\label{fig:tslengthbarplot}
\end{figure}

\vspace{2em}

\begin{figure}[H]
	\center
	\includegraphics[height=0.33\textheight]{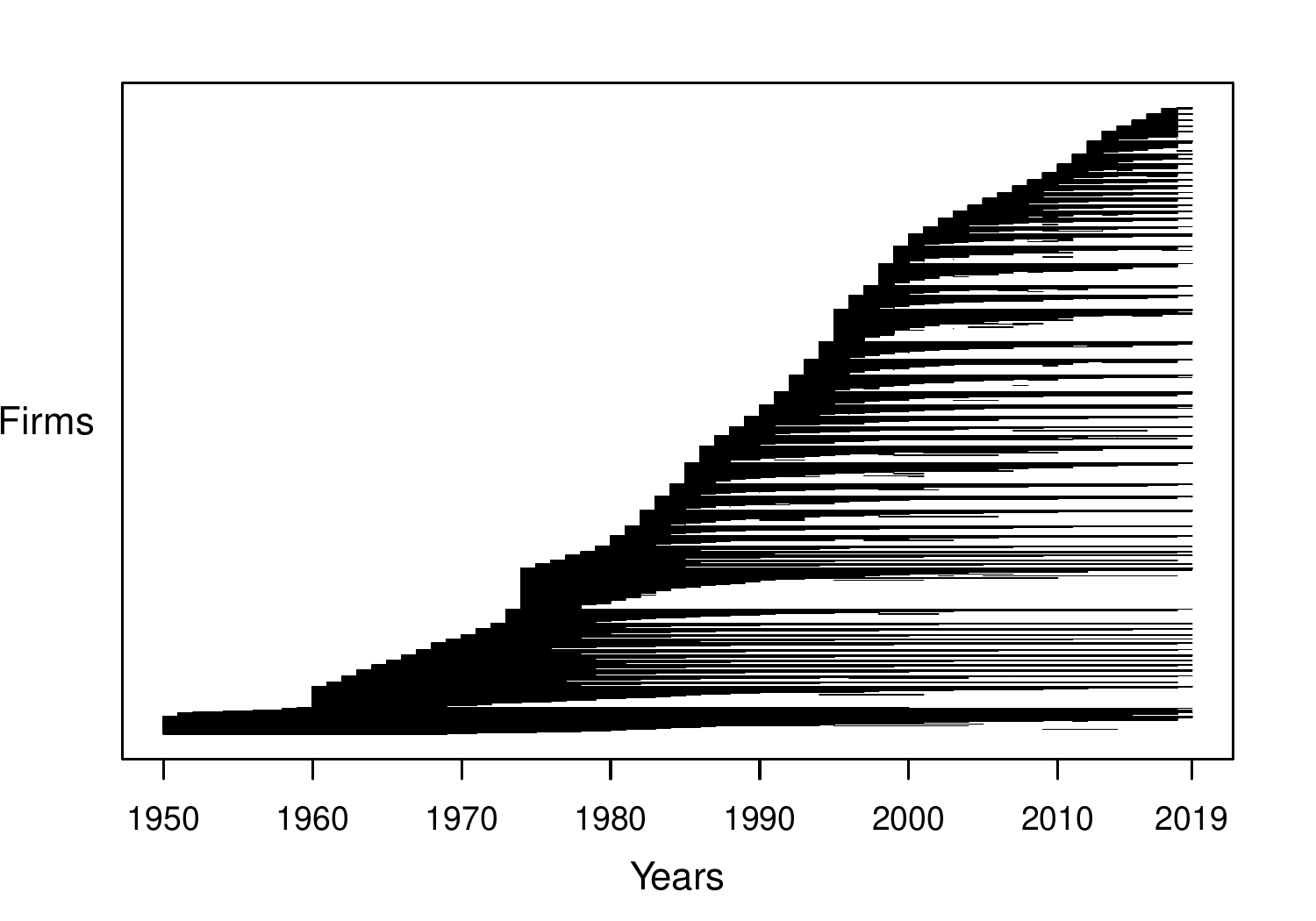}
	\caption{The figure displays the time series properties of the firms. Each of the 21,808 firms is represented by one horizontal line and these are ordered from bottom to top according to three criteria: 1. the first year of appearance in the data set, 2. the number of observations of the firm, 3. the number of consecutive observations of the firm.}
	\label{fig:tslength}
\end{figure}
\end{minipage}

\begin{landscape}
\begin{table}
    \hspace{-4em}
    \begin{minipage}[c]{\textwidth + 6em}
    \small
    \centering
    \setcapwidth[l]{\textwidth + 21em}
    \caption{\small Summary of predictor variables. CAGR is the compound annual growth rate, op. mar. C$\Delta$ is the compound operating margin delta, EBIT is the earning before interest and taxes, market cap. is the market capitalization, pp is percentage points and qu. is quantile. The summary on major and industry groups covers the group sizes.}
    \label{tab:predictors}
    \begin{tabular}{ll|rrrrrr|r}
         Predictor & Description & 2.5\% qu. & 25\% qu. & Median & Mean & 75\% qu. & 97.5\% qu. & Missings \\
        \midrule
        total assets & in million USD & 0.27 & 11.82 & 62.31 & 877.65 & 337.77 & 6767.24 & 2714 \\
        operating margin & EBIT divided by sales (in \%) & -827.80 & -1.19 & 6.01 & -402.68 & 12.27 & 34.49 & 18532 \\
        sales & in million USD & 0.00 & 10.67 & 67.60 & 721.10 & 337.74 & 5345.51 & 0 \\
        shareholder equity & total assets minus total liabilities (in million USD) & -9.65 & 3.58 & 24.00 & 319.76 & 128.97 & 2478.79 & 19811 \\
      major group & first two digits of SIC, 63 groups & 10 & 895 & 2646 & 4819.49 & 5295 & 25617 & 0 \\
      industry group & first three digits of SIC, 250 groups & 38 & 283 & 622 & 1214.51 & 1248 & 6793 & 0 \\
        $\beta$ & slope of regressing daily return on market return & -0.28 & 0.37 & 0.77 & 0.83 & 1.21 & 2.31 & 97469 \\
        price-to-book ratio & market cap. divided by shareholder equity & -6.00 & 0.59 & 1.34 & 2.65 & 2.57 & 11.70 & 100318 \\
        price-to-earnings ratio & market cap. divided by net income & -70.39 & -3.45 & 8.34 & 11.24 & 17.69 & 104.99 & 98786 \\
        past 1-year sales CAGR & sales growth rate in past year (in \%) & -100 & -5.39 & 4.93 & 115.70 & 19.24 & 1465000 & 31591 \\
        past 2-year sales CAGR & compound sales growth rate in past 2 years (in \%) & -100 & -4.18 & 4.55 & 17.07 & 16.33 & 19090 & 52164 \\
        past 3-year sales CAGR & compound sales growth rate in past 3 years (in \%) & -100 & -3.31 & 4.32 & 10.41 & 14.51 & 3862 & 71103 \\
        past 4-year sales CAGR & compound sales growth rate in past 4 years (in \%) & -100 & -2.71 & 4.21 & 7.90 & 13.17 & 1794 & 88572 \\
        past 5-year sales CAGR & compound sales growth rate in past 5 years (in \%) & -100 & -2.22 & 4.13 & 6.52 & 12.23 & 1019 & 104702 \\
        past 6-year sales CAGR & compound sales growth rate in past 6 years (in \%) & -100 & -1.87 & 4.05 & 5.62 & 11.44 & 609.50 & 119372 \\
        past 7-year sales CAGR & compound sales growth rate in past 7 years (in \%) & -100 & -1.55 & 4 & 5.02 & 10.82 & 435.80 & 132772 \\
        past 8-year sales CAGR & compound sales growth rate in past 8 years (in \%) & -100 & -1.29 & 3.98 & 4.59 & 10.38 & 333.90 & 145044 \\
        past 9-year sales CAGR & compound sales growth rate in past 9 years (in \%) & -100 & -1.06 & 3.95 & 4.28 & 9.97 & 277.10 & 156300 \\
        past 10-year sales CAGR & compound sales growth rate in past 10 years (in \%) & -100 & -0.87 & 3.91 & 4.03 & 9.58 & 205.30 & 166682 \\
        1-year op. mar. C$\Delta$ & difference to op. mar. 1 year ago (in pp) & -2824000 & -2.73 & 0.04 & -10.15 & 2.57 & 2823000 & 41527 \\
        2-year op. mar. C$\Delta$ & difference to op. mar. 2 years ago (in pp) & -1412000 & -1.96 & -0.03 & -11.85 & 1.71 & 681300 & 62660 \\
        3-year op. mar. C$\Delta$ & difference to op. mar. 3 years ago (in pp) & -374800 & -1.54 & -0.07 & 4.04 & 1.26 & 951200 & 81829 \\
        4-year op. mar. C$\Delta$ & difference to op. mar. 4 years ago (in pp) & -326200 & -1.27 & -0.08 & 3.89 & 1.00 & 691100 & 99288 \\
        5-year op. mar. C$\Delta$ & difference to op. mar. 5 years ago (in pp) & -260800 & -1.09 & -0.08 & 3.19 & 0.82 & 523200 & 115291 \\
        6-year op. mar. C$\Delta$ & difference to op. mar. 6 years ago (in pp) & -217300 & -0.95 & -0.09 & 0.42 & 0.69 & 204400 & 129585 \\
        7-year op. mar. C$\Delta$ & difference to op. mar. 7 years ago (in pp) & -107800 & -0.84 & -0.09 & 3.81 & 0.60 & 185700 & 142583 \\
        8-year op. mar. C$\Delta$ & difference to op. mar. 8 years ago (in pp) & -89290 & -0.76 & -0.08 & 2.25 & 0.53 & 190800 & 154449 \\
        9-year op. mar. C$\Delta$ & difference to op. mar. 9 years ago (in pp) & -81610 & -0.69 & -0.08 & 3.21 & 0.46 & 335300 & 165288 \\
        10-year op. mar. C$\Delta$ & difference to op. mar. 10 years ago (in pp) & -75350 & -0.64 & -0.08 & 3.44 & 0.41 & 301700 & 175265 \\
    \end{tabular}
    \end{minipage}
    \hspace{-6em}
\end{table}
\end{landscape}

\begin{minipage}[c][\textheight][c]{\textwidth}
\begin{table}[H]
    \center
    \caption{Compound annual sales growth rates for the whole data set. Mean and standard deviation are 2.5\% trimmed on both tails, the respective quantiles are contained in the table.}
    \label{tab:cagr-fulluniverse}
    \begin{tabular}{l|rrrr}
        Full Universe & \multicolumn{4}{c}{Base Rates}\\
        \midrule
        CAGR (\%) & 1-Yr & 3-Yr & 5-Yr & 10-Yr \\
        \midrule
        $\leq$ -25 & 8.70 & 5.44 & 4.00 & 2.38 \\
        ]-25,-20] & 2.19 & 1.69 & 1.28 & 0.68 \\
        ]-20,-15] & 3.18 & 2.65 & 2.13 & 1.37 \\
        ]-15,-10] & 4.53 & 4.27 & 3.71 & 2.68 \\
        ]-10,-5] & 7.06 & 7.28 & 7.11 & 6.12 \\
        ]-5,0] & 10.92 & 13.20 & 14.29 & 15.64 \\
        ]0,5] & 13.59 & 17.82 & 21.17 & 27.25 \\
        ]5,10] & 11.65 & 14.33 & 16.34 & 20.09 \\
        ]10,15] & 8.24 & 9.06 & 9.70 & 9.95 \\
        ]15,20] & 5.65 & 5.86 & 5.77 & 5.38 \\
        ]20,25] & 4.08 & 3.95 & 3.61 & 2.92 \\
        ]25,30] & 3.05 & 2.71 & 2.54 & 1.76 \\
        ]30,35] & 2.31 & 2.04 & 1.73 & 1.14 \\
        ]35,40] & 1.78 & 1.54 & 1.26 & 0.69 \\
        ]40,45] & 1.46 & 1.17 & 0.93 & 0.48 \\
        $>$ 45 & 11.58 & 6.99 & 4.42 & 1.46 \\
        \midrule
        mean & 10.62 & 7.01 & 5.75 & 4.62 \\
        \midrule
        median & 4.93 & 4.32 & 4.13 & 3.91 \\
        \midrule
        std & 32.30 & 19.08 & 14.21 & 9.20 \\
        \midrule
				$q_{0.025}$ & -60.01 & -44.75 & -36.52 & -23.91 \\
        \midrule
				$q_{0.975}$ & 206.31 & 95.19 & 62.75 & 35.85 \\
        \midrule
    \end{tabular}
\end{table}
\end{minipage}

\begin{minipage}[c][\textheight][c]{\textwidth}
\begin{figure}[H]
	\center
	\includegraphics[height=0.33\textheight]{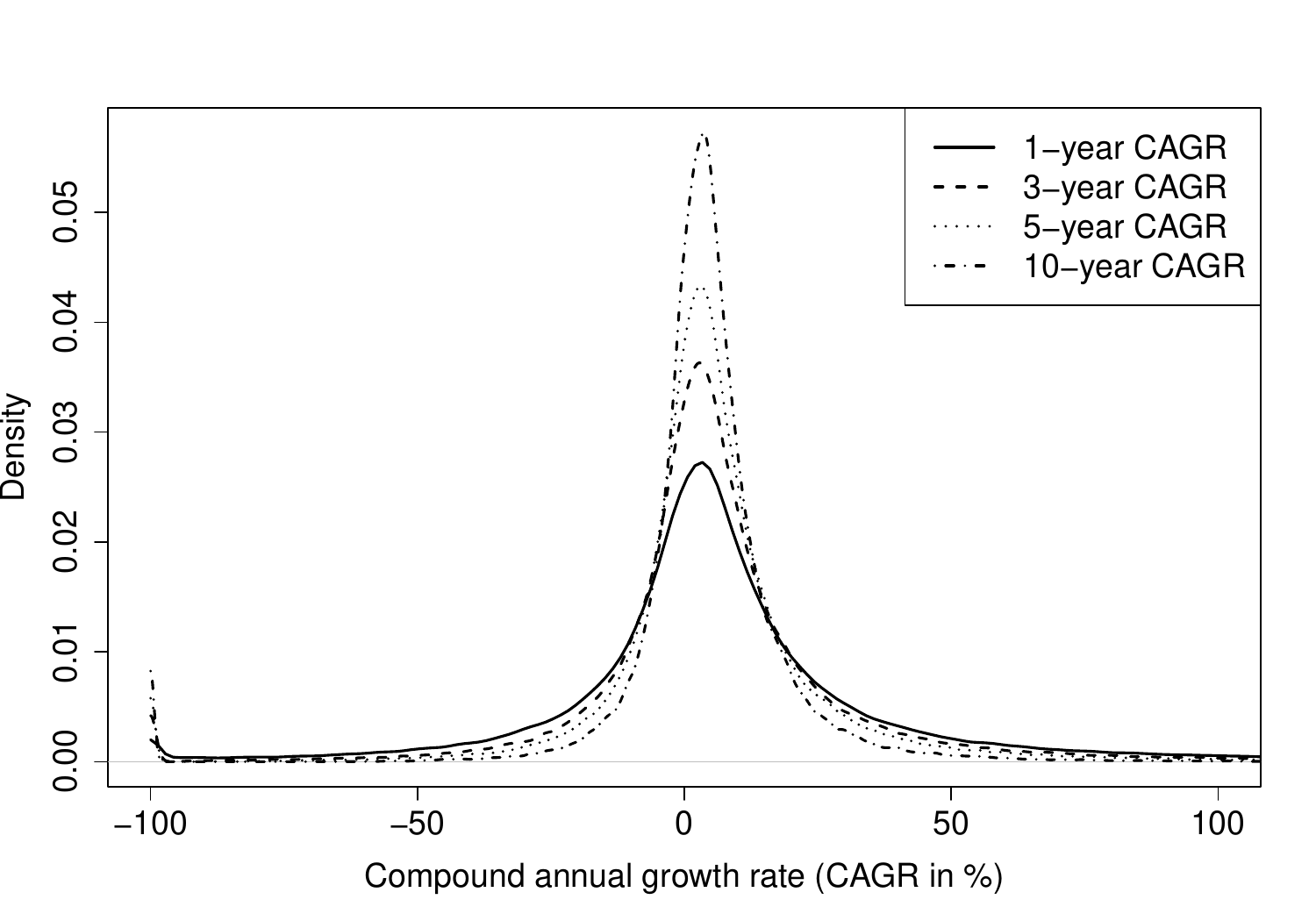}
	\caption{Estimated densities of compound annual sales growth for horizons one, three, five and 10 years. For density estimation on support $[-100,\infty )$ we used the Gaussian kernel with Silverman's rule of thumb as bandwidth.}
	\label{fig:cagr-dens}
\end{figure}

\vspace{7em}

\begin{table}[H]
	\center
	\caption{(Hyper) Parameters}
	\label{tab:parameters}
	\begin{tabular}{l|l}
		Name & Description \\
		\midrule
		predictor variable & see table \ref{tab:predictors} \\
		class size & relative size $\in \{0.050, 0.025, 0.010\}$ \\
		window & number of past years $\in \{5, 10, 20, 30\}$
	\end{tabular}
\end{table}
\end{minipage}

\newpage
\begin{minipage}[c][\textheight][c]{\textwidth}
\begin{table}[H]
    \center
    \caption{Comparison of predictor variables for forecasting one-year ahead sales growth}
    \label{tab:result1yearsalesgr}
    \makebox[\textwidth][c]{\begin{tabular}{lll|rrr}
        Predictor Variable & Window & Size & $\Delta_{\text{quantiles}}$ (rank) & KS (rank) & CvM (rank) \\
        \midrule
        six-year operating margin $\Delta$ & 30 & 0.025 & 0.0155 (1) & 1.874 (4) & 0.8265 (3) \\
        seven-year operating margin $\Delta$ & 30 & 0.025 & 0.0157 (2) & 2.1986 (10) & 1.0815 (8) \\
        six-year operating margin $\Delta$ & 30 & 0.01 & 0.0161 (3) & 2.4469 (14) & 1.3149 (13) \\
        operating margin & 10 & 0.05 & 0.0279 (24) & 4.1606 (50) & 6.1461 (74) \\
        operating margin & 5 & 0.05 & 0.0303 (26) & 4.4720 (74) & 4.8603 (43) \\
        sales (Mauboussin) & 5 & -- & 0.0516 (125) & 6.3825 (213) & 12.7518 (199) \\
        sales & 5 & 0.05 & 0.0524 (133) & 6.3939 (214) & 13.4453 (212) \\
        major group & 5 & -- & 0.0653 (201) & 8.6576 (274) & 22.5482 (256) \\
        industry group & 5 & -- & 0.0935 (295) & 10.7868 (302) & 36.6514 (291) \\
    \end{tabular}}
\end{table}

\vspace{10em}

\begin{table}[H]
    \center
    \caption{Comparison of predictor variables for forecasting three-year ahead sales growth}
    \label{tab:result3yearsalesgr}
    \makebox[\textwidth][c]{\begin{tabular}{lll|rrr}
        Predictor Variable & Window & Size & $\Delta_{\text{quantiles}}$ (rank) & KS (rank) & CvM (rank) \\
        \midrule
        seven-year operating margin $\Delta$ & 30 & 0.025 & 0.0286 (1) & 3.2227 (8) & 2.8868 (10) \\
        eight-year operating margin $\Delta$ & 30 & 0.025 & 0.0301 (2) & 1.9903 (2) & 1.0989 (1) \\
        eight-year operating margin $\Delta$ & 30 & 0.01 & 0.0302 (3) & 1.9878 (1) & 1.2532 (4) \\
        operating margin & 30 & 0.01 & 0.0598 (29) & 6.9177 (65) & 16.7632 (63) \\
        operating margin & 5 & 0.05 & 0.0697 (38) & 10.4675 (160) & 33.6971 (119) \\
        operating margin & 10 & 0.05 & 0.0877 (73) & 11.8366 (200) & 55.6297 (200) \\
        sales (Mauboussin) & 5 & -- & 0.1028 (143) & 13.4856 (247) & 61.3185 (211) \\
        sales & 5 & 0.05 & 0.1057 (155) & 13.8816 (253) & 63.7592 (213) \\
        major group & 5 & -- & 0.1423 (274) & 17.9423 (311) & 106.9768 (292) \\
        industry group & 30 & -- & 0.1863 (309) & 16.9141 (302) & 117.9496 (302) \\
    \end{tabular}}
\end{table}
\end{minipage}

\begin{minipage}[c][\textheight][c]{\textwidth}
\begin{table}[H]
    \center
    \caption{Comparison of predictor variables for forecasting five-year ahead sales growth}
    \label{tab:result5yearsalesgr}
    \makebox[\textwidth][c]{\begin{tabular}{lll|rrr}
        Predictor Variable & Window & Size & $\Delta_{\text{quantiles}}$ (rank) & KS (rank) & CvM (rank) \\
        \midrule
        10-year operating margin $\Delta$ & 30 & 0.01 & 0.0312 (1) & 2.204 (3) & 1.3081 (2) \\
        10-year operating margin $\Delta$ & 30 & 0.025 & 0.0341 (2) & 1.7507 (1) & 0.9922 (1) \\
        six-year operating margin $\Delta$ & 30 & 0.01 & 0.0361 (3) & 2.4614 (6) & 2.0039 (9) \\
        operating margin & 30 & 0.01 & 0.0851 (37) & 9.4868 (89) & 32.0685 (84) \\
        price-to-earnings ratio & 5 & 0.05 & 0.1096 (55) & 9.2194 (88) & 41.3370 (93) \\
        price-to-earnings ratio & 10 & 0.025 & 0.1485 (128) & 12.5293 (133) & 79.8237 (152) \\
        sales (Mauboussin) & 5 & -- & 0.1600 (170) & 19.0380 (277) & 137.3941 (261) \\
        sales & 5 & 0.05 & 0.1650 (187) & 19.5103 (279) & 147.1779 (269) \\
        major group & 30 & -- & 0.2136 (289) & 16.7058 (243) & 106.9918 (231) \\
        industry group & 30 & -- & 0.2179 (296) & 17.6483 (261) & 127.3253 (255) \\
    \end{tabular}}
\end{table}

\vspace{9em}

\begin{table}[H]
    \center
    \caption{Comparison of predictor variables for forecasting 10-year ahead sales growth}
    \label{tab:result10yearsalesgr}
    \makebox[\textwidth][c]{\begin{tabular}{lll|rrr}
        Predictor Variable & Window & Size & $\Delta_{\text{quantiles}}$ (rank) & KS (rank) & CvM (rank) \\
        \midrule
        six-year operating margin $\Delta$ & 30 & 0.025 & 0.0432 (1) & 3.7904 (5) & 4.1498 (5) \\
        seven-year operating margin $\Delta$ & 30 & 0.025 & 0.0456 (2) & 3.5849 (3) & 3.8386 (2) \\
        five-year operating margin $\Delta$ & 30 & 0.025 & 0.0478 (3) & 4.0971 (15) & 5.0842 (9) \\
        operating margin & 30 & 0.01 & 0.1112 (36) & 7.4423 (80) & 20.6308 (88) \\
        10-year operating margin $\Delta$ & 10 & 0.025 & 0.2033 (113) & 8.5930 (103) & 31.8499 (106) \\
        sales & 30 & 0.01 & 0.2099 (115) & 10.0584 (112) & 42.9767 (118) \\
        sales (Mauboussin) & 30 & -- & 0.2270 (128) & 11.2416 (130) & 50.6546 (128) \\
        major group & 30 & -- & 0.2561 (146) & 12.0198 (131) & 61.4773 (134) \\
        price-to-earnings ratio & 5 & 0.01 & 0.2842 (168) & 17.4874 (183) & 136.6147 (192) \\
        industry group & 30 & -- & 0.2859 (169) & 13.4787 (141) & 75.4007 (145) \\
    \end{tabular}}
\end{table}
\end{minipage}

\begin{minipage}[c][\textheight][c]{\textwidth}
\begin{table}[H]
    \center
    \caption{Influence of the best predictor variables on median, mean and standard deviation of the reference classes for forecasting compound sales growth for different forecasting horizons. Mean and standard deviation are 2.5\% trimmed on both tails. op.mar $\Delta_l$ stands for l-year operating margin delta and is measured in percentage points per year.}
    \label{tab:pred-infl}
    \begin{tabular}{l|l|rrr||l|rrr}
    	 & \multicolumn{4}{c}{one-year forecast horizon} & \multicolumn{4}{c}{three-year forecast horizon} \\
        qu. & op.mar. $\Delta_{6}$ & median & mean & std & op.mar. $\Delta_{7}$ & median & mean & std \\
        \midrule
        10\% & -3.50 & -0.04 & 1.65 & 26.72 & -2.74 & 0.43 & 0.43 & 17.66 \\
        20\% & -1.44 & 0.66 & 1.28 & 17.58 & -1.19 & 0.81 & 0.86 & 11.83  \\
        30\% & -0.74 & 1.39 & 1.97 & 14.62 & -0.62 & 1.68 & 2.12 & 10.17  \\
        40\% & -0.33 & 2.39 & 3.04 & 13.98 &  -0.28 & 1.93 & 2.20 & 10.03 \\
        50\% & -0.03 & 3.40 & 4.64 & 12.47 & -0.02 & 2.55 & 3.06 & 9.57 \\
        60\% & 0.27 & 3.16 & 4.18 & 12.62 & 0.23 & 2.48 & 3.01 & 9.43 \\
        70\% & 0.68 & 3.77 & 4.93 & 14.07 & 0.58 & 2.92 & 3.73 & 9.73 \\
        80\% & 1.44 & 3.66 & 5.23 & 17.69 & 1.20 & 3.34 & 4.34 & 12.28 \\
        90\% & 4.48 & 4.67 & 7.36 & 28.57 & 3.51 & 4.16 & 5.31 & 17.88 \\
        \midrule
        \midrule
         & \multicolumn{4}{c}{five-year forecast horizon} & \multicolumn{4}{c}{10-year forecast horizon} \\
         qu. & op.mar. $\Delta_{10}$ & median & mean & std & op.mar. $\Delta_{6}$  & median & mean & std \\
        \midrule
        10\% & -1.74 & 0.40 & 0.59 & 11.84 & -2.68 & 1.49 & 1.32 & 10.82  \\
        20\% & -0.83 & 1.27 & 1.20 & 9.58 & -1.19 & 2.37 & 2.68 & 6.75  \\
        30\% & -0.47 & 2.31 & 2.48 & 8.55 & -0.63 & 1.58 & 1.78 & 6.36\\
        40\% & -0.22 & 1.44 & 1.56 & 8.57 & -0.27 & 2.46 & 2.68 & 6.25 \\
        50\% & -0.04 & 2.04 & 2.15 & 7.14 & 0.00 & 2.73 & 2.59 & 5.96 \\
        60\% & 0.14 & 3.24 & 3.40 & 8.44 & 0.27 & 3.02 & 3.42 & 6.16 \\
        70\% & 0.37 & 1.87 & 2.49 & 7.96 & 0.63 & 2.96 & 3.28 & 6.49 \\
        80\% & 0.75 & 2.69 & 3.21 & 8.81 & 1.27 & 3.04 & 3.58 & 7.19 \\
        90\% & 2.05 & 3.15 & 4.55 & 13.36 & 3.59 & 4.82 & 4.97 & 10.55 \\
    \end{tabular}
\end{table}
\end{minipage}

\begin{minipage}[c][\textheight][c]{\textwidth}
\begin{figure}[H]
	\center
	\includegraphics[height=0.33\textheight]{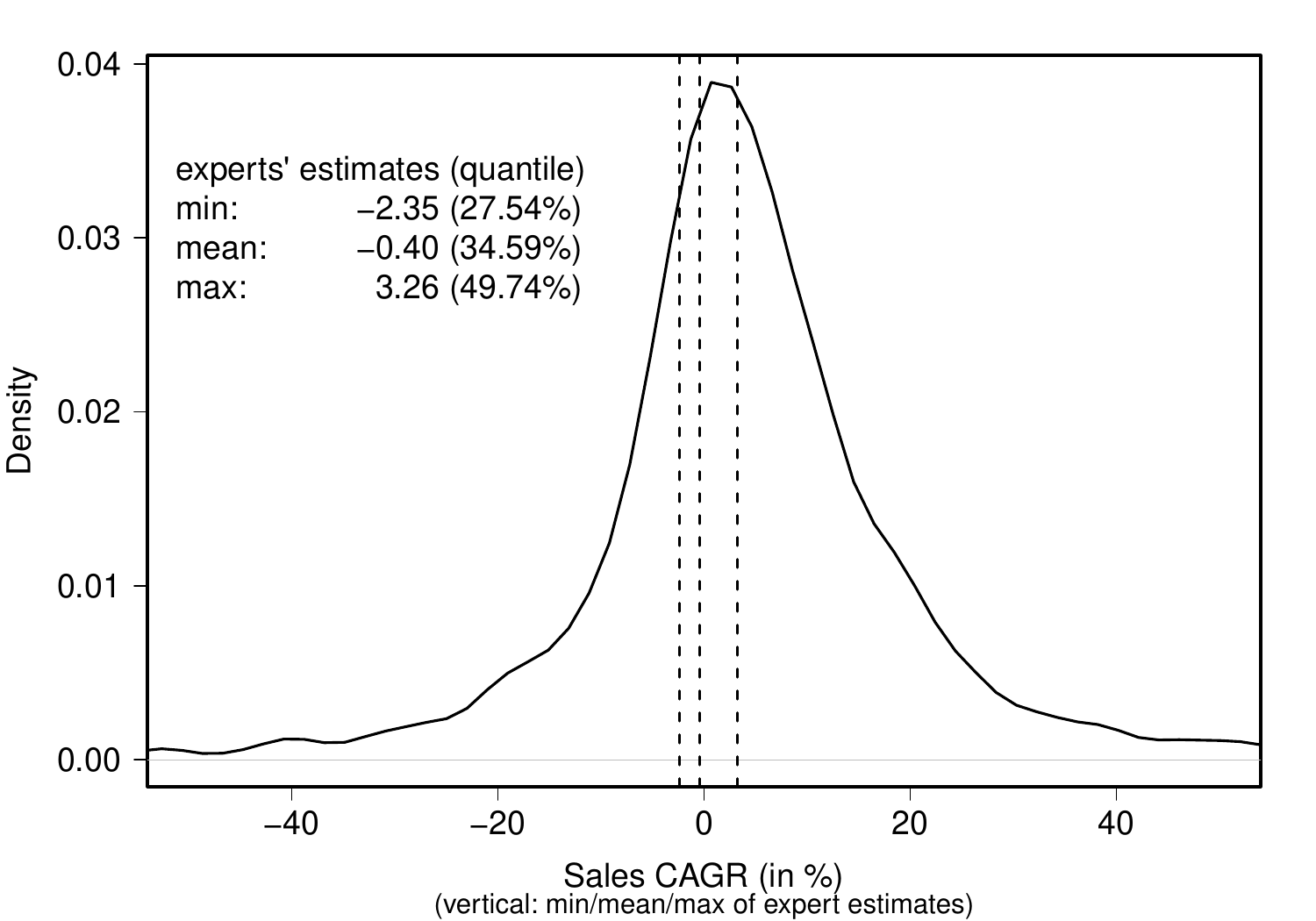}
	\caption{Forecasted density of one-year sales growth for 3M based on six-year operating margin delta (1.77 percentage points) and with hyper parameters window = 30 and size = 0.025 compared to experts' estimates. For density estimation on support $[-100,\infty )$ we used the Gaussian kernel with Silverman's rule of thumb as bandwidth.}
	\label{fig:example-dreim-1}
\end{figure}

\vspace{2em}

\begin{figure}[H]
	\center
	\includegraphics[height=0.33\textheight]{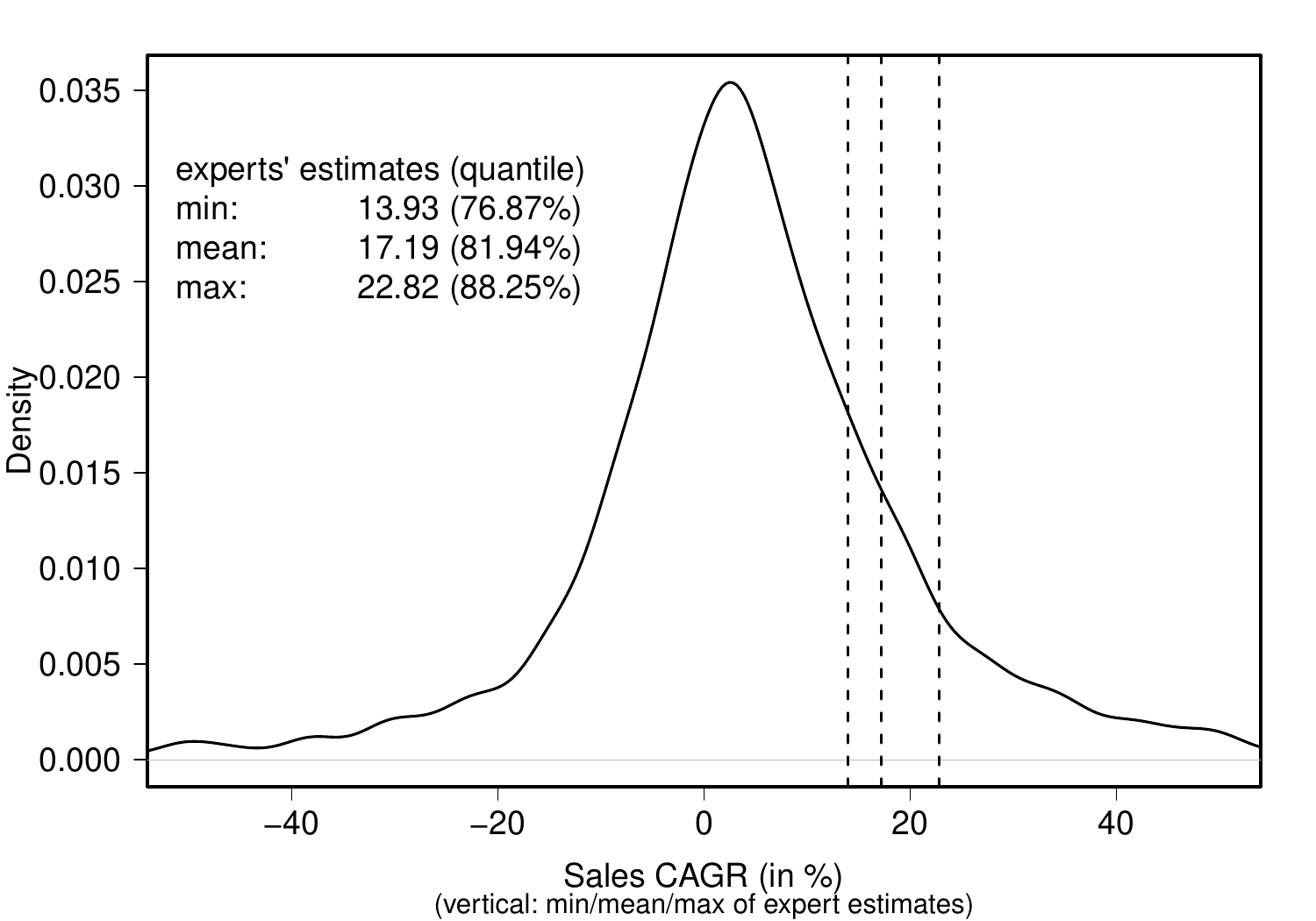}
	\caption{Forecasted density of one-year sales growth for Amazon based on six-year operating margin delta (4.16 percentage points) and with hyper parameters window = 30 and size = 0.025 compared to experts' estimates. For density estimation on support $[-100,\infty )$ we used the Gaussian kernel with Silverman's rule of thumb as bandwidth.}
	\label{fig:example-amazon-1}
\end{figure}
\end{minipage}

\begin{minipage}[c][\textheight][c]{\textwidth}
\begin{table}[H]
    \center
    \caption{Comparison of base rates for 3M based on reference classes of our approach using the respective best predictor and hyper parameters and of \cite{maubcal15}. Mean and standard deviation are 2.5\% trimmed on both tails.}
    \label{tab:refclass_dreim}
        \begin{tabular}{l|rr|rr|rr|rr}
        3M & \multicolumn{8}{c}{Base Rates}\\
        \midrule
        CAGR (\%) & 1-Yr & 1-Yr MC & 3-Yr & 3-Yr MC & 5-Yr & 5-Yr MC & 10-Yr & 10-Yr MC \\
        \midrule
        $\leq$ -25 & 4.13 & 4.64 & 2.12 & 1.53 & 1.16 & 0.97 & 0.57 & 0.41 \\
        ]-25,-20] & 1.50 & 1.71 & 1.77 & 2.39 & 0.66 & 0.83 & 0.43 & 0.26 \\
        ]-20,-15] & 2.71 & 2.92 & 1.58 & 4.11 & 2.64 & 2.07 & 1.42 & 1.31 \\
        ]-15,-10] & 4.01 & 4.42 & 3.89 & 5.40 & 3.31 & 4.77 & 2.83 & 2.90 \\
        ]-10,-5] & 7.86 & 8.72 & 8.87 & 10.67 & 8.43 & 11.20 & 6.02 & 9.65 \\
        ]-5,0] & 16.16 & 19.37 & 18.04 & 26.13 & 18.51 & 27.37 & 17.08 & 27.77 \\
        ]0,5] & 20.17 & 24.17 & 24.25 & 26.07 & 32.23 & 29.72 & 35.79 & 35.65 \\
        ]5,10] & 14.95 & 15.95 & 15.57 & 13.62 & 15.70 & 15.41 & 20.55 & 15.72 \\
        ]10,15] & 9.96 & 6.46 & 9.41 & 5.09 & 8.43 & 3.87 & 8.79 & 4.11 \\
        ]15,20] & 6.36 & 3.48 & 5.47 & 2.21 & 4.63 & 2.07 & 3.97 & 1.32 \\
        ]20,25] & 3.73 & 2.48 & 3.65 & 1.10 & 2.64 & 0.76 & 1.63 & 0.51 \\
        ]25,30] & 2.15 & 1.55 & 1.53 & 0.67 & 0.66 & 0.41 & 0.57 & 0.27 \\
        ]30,35] & 1.58 & 1.16 & 1.43 & 0.18 & 0.99 & 0.35 & 0.14 & 0.09 \\
        ]35,40] & 1.05 & 0.77 & 0.59 & 0.18 & 0.00 & 0.14 & 0.14 & 0.02 \\
        ]40,45] & 0.45 & 0.72 & 0.69 & 0.12 & 0.00 & 0.00 & 0.07 & 0.00 \\
        $>$ 45 & 3.24 & 1.49 & 1.13 & 0.49 & 0.00 & 0.07 & 0.00 & 0.00 \\
        \midrule
        mean & 4.30 & 1.59 & 3.54 & -0.45 & 2.57 & 0.29 & 3.18 & 0.89 \\
        \midrule
        median & 3.33 & 1.73 & 2.53 & -0.04 & 2.38 & 0.31 & 3.02 & 0.92 \\
        \midrule
        std & 12.89 & 11.31 & 10.09 & 7.80 & 7.66 & 6.32 & 6.30 & 5.13 \\
        \midrule
    \end{tabular}
\end{table}
\end{minipage}

\begin{minipage}[c][\textheight][c]{\textwidth}
\begin{table}[H]
    \center
    \caption{Comparison of base rates for Amazon based on reference classes of our approach using the respective best predictor and hyper parameters and of \cite{maubcal15}. Mean and standard deviation are 2.5\% trimmed on both tails.}
    \label{tab:refclass_amazon}
    \begin{tabular}{l|rr|rr|rr|rr}
        Amazon & \multicolumn{8}{c}{Base Rates}\\
        \midrule
        CAGR (\%) & 1-Yr & 1-Yr MC & 3-Yr & 3-Yr MC & 5-Yr & 5-Yr MC & 10-Yr & 10-Yr MC \\
        \midrule
        $\leq$ -25 & 4.37 & 3.31 & 1.72 & 1.23 & 1.16 & 2.08 & 1.06 & 0.35 \\
        ]-25,-20] & 1.74 & 0.55 & 1.77 & 3.68 & 0.83 & 0.00 & 0.57 & 0.71 \\
        ]-20,-15] & 2.39 & 3.87 & 1.72 & 4.29 & 2.31 & 4.17 & 1.63 & 2.36 \\
        ]-15,-10] & 4.50 & 2.76 & 3.55 & 4.29 & 2.81 & 3.47 & 2.20 & 2.60 \\
        ]-10,-5] & 8.95 & 8.29 & 7.93 & 11.04 & 8.60 & 15.97 & 6.87 & 10.64 \\
        ]-5,0] & 14.54 & 17.68 & 19.02 & 19.02 & 18.35 & 16.67 & 20.84 & 30.02 \\
        ]0,5] & 18.47 & 26.52 & 22.77 & 28.22 & 33.06 & 32.64 & 32.67 & 33.33 \\
        ]5,10] & 13.69 & 16.02 & 18.43 & 17.79 & 15.87 & 19.44 & 20.77 & 16.31 \\
        ]10,15] & 10.04 & 6.63 & 9.96 & 5.52 & 9.09 & 4.17 & 6.52 & 2.84 \\
        ]15,20] & 6.97 & 4.42 & 5.32 & 3.07 & 2.98 & 0.00 & 4.46 & 0.71 \\
        ]20,25] & 3.93 & 5.52 & 2.37 & 1.23 & 2.31 & 0.69 & 1.35 & 0.12 \\
        ]25,30] & 2.59 & 1.66 & 1.38 & 0.61 & 1.32 & 0.69 & 0.50 & 0.00 \\
        ]30,35] & 1.94 & 1.10 & 1.28 & 0.00 & 0.50 & 0.00 & 0.50 & 0.00 \\
        ]35,40] & 1.26 & 1.10 & 0.84 & 0.00 & 0.50 & 0.00 & 0.00 & 0.00 \\
        ]40,45] & 1.09 & 0.55 & 0.34 & 0.00 & 0.17 & 0.00 & 0.00 & 0.00 \\
        $>$ 45 & 3.52 & 0.00 & 1.58 & 0.00 & 0.17 & 0.00 & 0.07 & 0.00 \\
        \midrule
        mean & 4.93 & 2.75 & 3.72 & 0.00 & 2.55 & 0.03 & 2.65 & 0.16 \\
        \midrule
        median & 3.70 & 2.27 & 2.88 & 0.39 & 2.16 & 1.23 & 2.50 & 0.49 \\
        \midrule
        std & 14.11 & 10.73 & 9.74 & 8.23 & 7.62 & 7.01 & 6.46 & 5.15 \\
        \midrule
    \end{tabular}
\end{table}
\end{minipage}

\end{document}